\documentclass[sigplan]{acmart}
\makeatletter                   
\def\mdseries@tt{m}             
\makeatother                    
\acmConference[CGO 2020]{International Symposium on Code Generation and Optimization}{February 22--26, 2020}{\textit{to appear}}

\acmYear{}
\acmISBN{} 
\acmDOI{} 
\startPage{1}

\setcopyright{none} 

\usepackage{booktabs}   
\usepackage{subcaption} 
\usepackage{mathtools} 
\usepackage[draft=true]{minted} 

\usepackage{amsmath}
\usepackage{amsfonts}
\usepackage{tikz}

\newcommand*\circled[1]{\tikz[baseline=(char.base)]{
            \node[shape=circle,draw,inner sep=0.75pt] (char) {\small #1};}}

\DeclareMathOperator*{\argmaxA}{arg\,max}

\begin{document}

\title[NeuroVectorizer: End-to-End Vectorization with Deep RL]{NeuroVectorizer: End-to-End Vectorization with Deep Reinforcement Learning}         
\titlenote{Work done while Haj-Ali was at Intel Labs.}             

\author[Haj-Ali~\textit{et al.}]{Ameer~Haj-Ali$^{1}$\ \ \ \ Nesreen~K.~Ahmed$^2$\ \ \ \ Ted~Willke$^2$\ \ \ \ Yakun~Sophia~Shao$^1$\ \ \ \ Krste~Asanovic$^1$\ \ \ \ Ion~Stoica$^1$\\
$^1$University of California, Berkeley, $^2$Intel Labs \\
\{ameerh, ysshao, krste, istoica\}@berkeley.edu, \{nesreen.k.ahmed, ted.willke\}@intel.com}
\begin{abstract}
One of the key challenges arising when compilers vectorize loops for today's SIMD-compatible architectures is to decide if vectorization or interleaving is beneficial. 
Then, the compiler has to determine how many instructions to pack together and how many loop iterations to interleave.
Compilers are designed today to use fixed-cost models that are based on heuristics to make vectorization decisions on loops.
However, these models are unable to capture the data dependency, the computation graph, or the organization of instructions.
Alternatively, software engineers often hand-write the vectorization factors of every loop. 
This, however, places a huge burden on them, since it requires prior experience and significantly increases the development time. 

In this work, we explore a novel approach for handling loop vectorization and propose an end-to-end solution using deep reinforcement learning (RL). 
We conjecture that deep RL can capture different instructions, dependencies, and data structures to enable learning a sophisticated model that can better predict the actual performance cost and determine the optimal vectorization factors. 
We develop an end-to-end framework, from code to vectorization, that integrates deep RL in the LLVM compiler. Our proposed framework takes benchmark codes as input and extracts the loop codes. These loop codes are then fed to a loop embedding generator that learns an embedding for these loops. Finally, the learned embeddings are used as input to a Deep RL agent, which dynamically determines the vectorization factors for all the loops. We further extend our framework to support random search, decision trees, supervised neural networks, and nearest-neighbor search. 
We evaluate our approaches against the currently used LLVM vectorizer and loop polyhedral optimization techniques. Our experiments show $1.29\times-4.73\times$ performance speedup compared to baseline and only $3\%$ worse than the brute-force search on a wide range of benchmarks.

\end{abstract}




\maketitle

\section{Introduction}
\label{intro}
Modern computers typically have vector instructions that perform multiple basic operations simultaneously, such as Intel Advanced Vector Extensions (AVX)~\cite{lomont2011introduction}. Converting a computer program from a scalar implementation, which processes a single pair of operands at a time to a vector implementation, which performs a single operation on multiple data (SIMD) items at once is called \emph{vectorization}, and is critical to enhancing the performance of compute-intensive programs for modern computers.

Loops are among the most commonly vectorized parts of code. Loop vectorization is done by defining the vectorization factor (VF) and the interleaving factor (IF)~\cite{nuzman2006auto}. VF determines how many instructions to pack together from different iterations of the loop. IF determines the stride of the memory accesses of the packed instructions. IF allows vectorization to be performed on non-consecutive addresses, which are generally referred to as non-unit stride
accesses.

In most C and C++ compilers it is possible to use intrinsic pragmas or compiler passes to manually vectorize loops by setting the VF and IF.  However, manual vectorization is labor-intensive, error-prone, and results in code that is difficult to maintain and port.  Alternatively, several solutions for automatic vectorization and loop optimization have been proposed. The current vectorizer used in LLVM and proposed improvements~\cite{tian2016llvm,trifunovic2009polyhedral}, rely on linear and constant-cost models to predict the vectorization factors. Unfortunately, these cost models do not consider the computation graph and focus on estimating the cost of different instructions with predefined heuristics. Another commonly used approach is Polly~\cite{grosser2012polly}. Polly uses loop polyhedral analysis, which relies on an abstract mathematical representation, namely equations and matrices, to represent loops as polytopes. The polytope representation simplifies the implementation of loop optimizations, though to date the main optimizations in Polly are tiling and loop fusion to improve data locality.

Machine learning is yet another recent approach that has been proposed for automatic vectorization~\cite{stock2012using}. While this approach improves the cost models implemented by existing compilers, they use hand-engineered heuristics to extract features from the assembly code, such as arithmetic intensity. Unfortunately, these features are typically not sufficient to fully capture the code functionality. To overcome this challenge, \cite{cummins2017end} proposed an end-to-end solution that relies on deep supervised learning. However, supervised learning methods require labels to train. These labels are not always available and it can be time-consuming to find them. Furthermore, optimizing for multiple objectives with large search spaces can be challenging for supervised learning methods. 
 \begin{figure*}[!t]
     \centering
     \includegraphics[trim={30mm 0.5mm 42mm 9mm},clip,width=\textwidth]{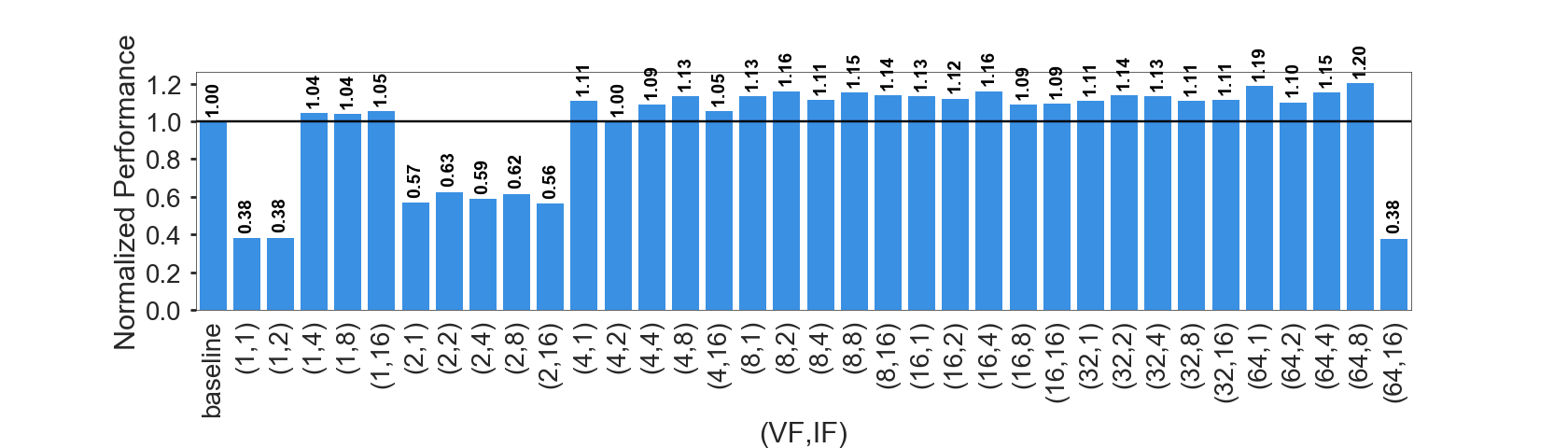}
     \caption{Performance of the dot product kernel for different VFs and IFs, normalized to the baseline cost model implemented in LLVM. The best VF and IF  corresponding to the baseline cost model are $(VF=4,IF=2)$}
     \label{fig:mot1}
 \end{figure*}

A human vectorization expert can determine the optimal vectorization factors, \textit{i.e.}, VF and IF for a specific hardware architecture by examining the computation graph, functionality, operations, and loop bodies in the text code. Similarly, in this paper, we use a code embedding generator that reads the text similar to a human expert, "understands" it and then generates an embedding that represents it. We use the generated embedding as an input to another neural network that can learn a mapping from this embedding to optimal vectorization factors similar to those learned by a human expert. This approach efficiently addresses the vectorization challenge end-to-end: from code to optimal factors, enabling the co-optimization of multiple objectives while preserving code correctness.

This paper makes the following contributions:
\begin{itemize}
    \item A comprehensive data set of more than 10,000 synthetic loop examples.
    \item An end-to-end deep reinforcement learning (RL)~\cite{sutton2018reinforcement} based auto-vectorization method.
    \item An extensible, open-source\footnote{\href{https://github.com/Intel-Academic/NeuroVectorizer}{https://github.com/Intel-Academic/NeuroVectorizer.}} framework that integrates learning code embedding with multiple machine learning methods to make vectorization predictions on loops. We explore using random search, supervised learning methods, \textit{i.e.}, nearest-neighbor search (NNS)~\cite{roussopoulos1995nearest}, decision trees~\cite{quinlan1986induction}, and supervised fully connected neural network (FCNNs), and contextual bandits based on deep RL. 
    \item Rigorous evaluations across different learning hyperparameters and benchmark suites to show the effectiveness of our approaches versus the currently used cost model, as well as Polly. Our results show $1.29\times-4.73\times$ average performance speedup and only $3\%$ worse than the brute-force solution.

\end{itemize}

The rest of the paper is organized as follows. In Section~\ref{motivation} motivation and background for using deep RL to automatically vectorize loops is given. The framework for automatic vectorization with deep RL is proposed in Section~\ref{framework} and evaluated on a wide range of benchmarks in Section~\ref{eval}. Future directions and related work are given in Sections~\ref{future} and \ref{related}, respectively. The paper is concluded in Section~\ref{conc}.

\section{Motivation and Background}
\label{motivation}
\subsection{Vectorization Characterization}
The vectorization is critical to enhancing the performance of compute-intensive workloads in modern computers. All the dedicated vector machines and modern CPUs that support vector instructions rely on vectorization to enhance the performance of such workloads.

Loops are among the most commonly vectorized parts of codes. Loop vectorization is done by setting the VF and the IF, which respectively determine the number of instructions to pack together and the stride. Appropriately setting the values of VF and IF for loops is cumbersome as it depends on many parameters, such as the instructions in the loop body, the stride, the underlying hardware architecture, the computations graph, and the functionality.

To understand this challenge and motivate this work we take a simple vector dot product kernel function:
\begin{minted}[fontsize=\footnotesize]{c}
 int vec[512] __attribute__((aligned(16)));
 __attribute__((noinline))
 int dot_product () {
     int sum = 0;
     for(int i = 0; i<512; i++){
         sum += vec[i]*vec[i];
     }
     return sum;
 }
\end{minted}
To eliminate noise and variance in results we run this kernel one million times and average the execution time. We run the kernel on 16 GB 2133 MHz LPDDR3 memory and 2.7 GHz (up to 4.5 GHz) Intel 4-Core i7-8559U~\cite{IntelInc2018}, which supports AVX. Figure~\ref{fig:mot1} shows the performance of this kernel after a brute-force search for different VFs and IFs normalized to the baseline cost model implemented in LLVM. The best VF and IF  corresponding to the baseline cost model are $(VF=4, IF=2)$. While the baseline improved the performance by $2.6\times$ when compared to the unvectorized code $(VF=1, IF=1)$, we can still see that $26$ out of $35$ possible factors improve over the baseline. This improvement is maximized by $(VF=64, IF=8)$ which achieves up to 20\% better performance than the baseline.

\begin{figure}[!t]
    \centering
    \includegraphics[trim={4mm 6.5mm 12.5mm 15mm}, clip, width=0.47\textwidth]{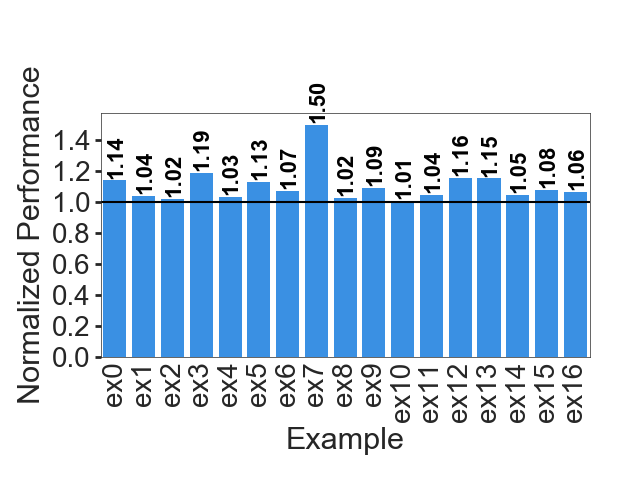}
    \caption{Performance of brute-force search of LLVM's vectorizer test suite, normalized to the baseline cost model implemented in LLVM.}
    \label{fig:mot2}
\end{figure}

\begin{figure*}[!t]
    \centering
    \includegraphics[width=0.8\textwidth]{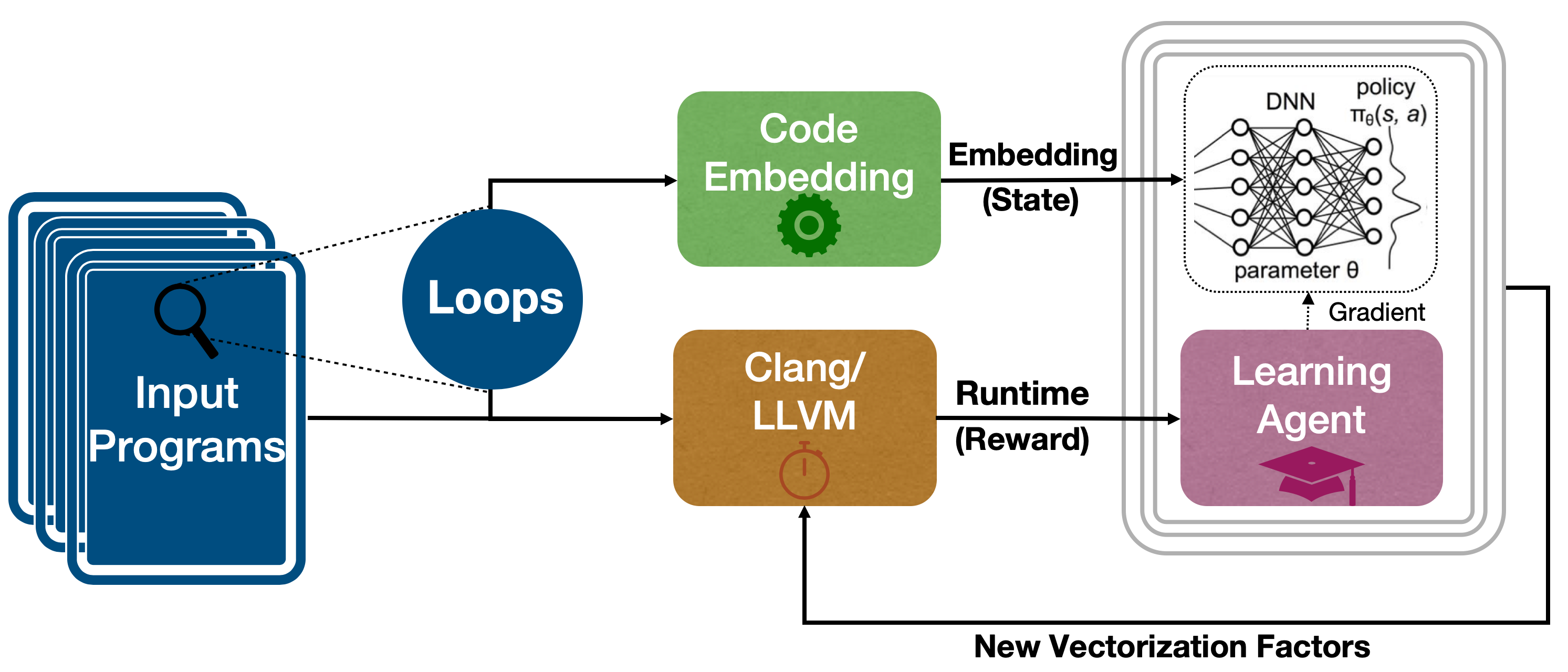}
    \caption{The proposed framework for automatic vectorization with deep RL. The programs are read to extract the loops. The loop texts are fed to the code embedding generator to generate an embedding. The embedding is fed to the RL agent. The RL agent learns a policy that maps this embedding to optimal vectorization factors by injecting compiler pragmas and compiling the programs with Clang/LLVM to gather the rewards: the execution time improvements.}
    \label{fig:framework}
\end{figure*}
\begin{figure*}
    \centering
    \includegraphics[width=\textwidth]{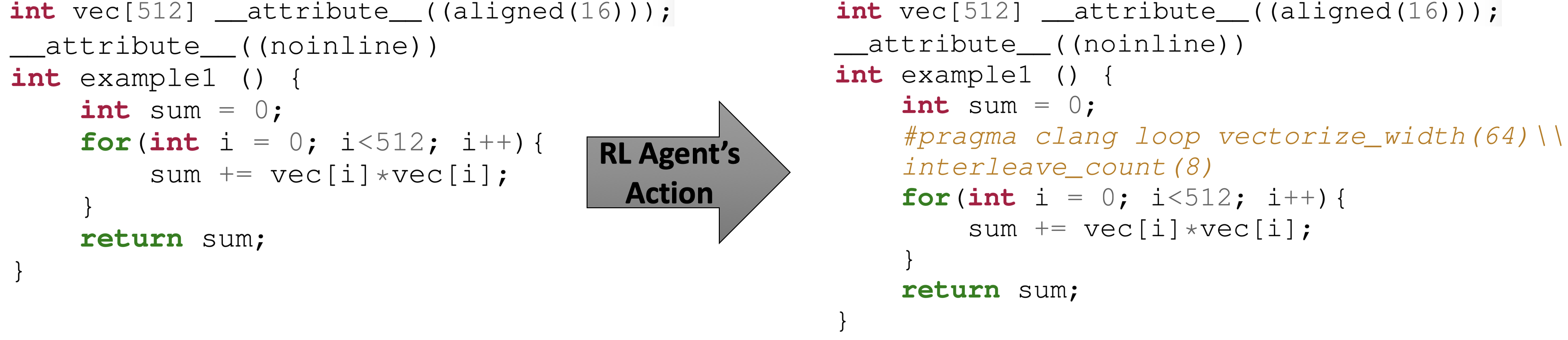}
    \caption{An example of the automatically injected VF and IF pragmas by the RL agent.}
    \label{fig:action}
\end{figure*}

To further motivate this work, we evaluate the vectorization test suite used in the LLVM base code\footnote{The test suite is available on: \href{https://github.com/llvm/llvm-test-suite/tree/master/SingleSource/UnitTests/Vectorizer}{https://github.com/llvm/llvm-test-suite/tree/master/SingleSource/UnitTests/Vectorizer}.}, which tests the cost model of the baseline vectorizer in LLVM. We run a brute-force search on all the possible VFs and IFs. The performance of the optimal vectorization normalized to the baseline is illustrated in Figure~\ref{fig:mot2}. In all the tests, the optimal vectorization performed better than the baseline. This performance gap increases with more complicated tests reaching up to $1.5\times$. These abstract results on simple tests show that there is room for improvement for the current baseline cost model.

\subsection{State-of-the-Art Auto-Vectorzation}

Most C and C++ compilers allow the users to manually determine the VF and the IF in their code. This, however, is error-prone, time-consuming and often not optimal. Thus, many works have been proposed in the past to address the automatic vectorization challenge. For example, Polly~\cite{grosser2012polly} uses an abstract mathematical representation based on integer polyhedra to analyze and optimize the memory access pattern of a program. Polly performs classical loop transformations, especially tiling and loop fusion to improve data-locality. These transformations also simplify vectorization decisions for the compiler. Accordingly, to date, the main optimizations in Polly are tiling and loop fusion to improve data locality.

Prior work~\cite{stock2012using} represented the code characteristics, by using hand-engineered heuristics extracted from the assembly code, such as arithmetic intensity and used it in conjunction with supervised learning to predict the vectorization factors. Unfortunately, these features are typically not sufficient to fully capture the code functionality~\cite{cummins2017synthesizing}. To overcome this challenge, \cite{cummins2017end} proposed an end-to-end solution that relies on deep supervised learning. However, supervised learning methods require labels to train and finding these labels can be time-consuming. Furthermore, optimizing for multiple objectives with large search spaces can be challenging for supervised learning methods. 

To appropriately set the VF and IF for the loops, it is necessary to fully learn the characteristics of the code and then use these characteristics to predict the optimal VF and IF. In other words, it is necessary to extract the loops from the code, characterize them, and use this characterization to predict the optimal factors. Therefore, we propose and develop a framework that accomplishes this goal by extracting the loops from the code, learning an embedding for these loops and learning a mapping from this embedding to the optimal VF and IF in an end-to-end fashion with RL. Unlike supervised learning methods, deep RL can be tuned to co-optimize multiple objectives and does not require a brute-force search and thus it can be more sample efficient.

\subsection{Deep Reinforcement Learning for Auto-Vectorization}
One of the promising machine learning approaches is RL, in which an agent learns by continually interacting with an environment~\cite{kaelbling1996reinforcement}. Using a neural network in conjunction with RL is called deep RL. Deep models allow RL algorithms to solve complex problems in an end-to-end fashion, handle unstructured environments, learn complex functions, or predict actions in states that have not been visited in the past. Deep RL is gaining wide interest recently due to its success in robotics, Atari gameplay, and superhuman capabilities~\cite{mnih2013playing,doya2000reinforcement,kober2013reinforcement,peters2003reinforcement,hajali2019deep}. Deep RL was the key technique behind defeating the human European champion in the game of Go, which has long been viewed as the most challenging of classic games for artificial intelligence~\cite{silver2016mastering}. 

In RL, the agent observes the state of the environment, and based on this state/observation takes an action. The ultimate goal is to compute a policy ($\pi^*$)--a mapping between the environment states and actions--that maximizes expected reward:
\begin{align}
\label{eq:MDP}
    \pi^* = \argmaxA_\pi \mathbb{E}_{\tau{\raise.05ex\hbox{$\scriptstyle\mathtt{\sim}$}}\pi(\tau)} \left[\tau \right].
\end{align}
where $\tau$ is a sequence of states and actions that define a single episode.

If the number of steps the RL agent has to take before the environment terminates is one, the problem is called Contextual Bandits. In Contextual Bandits the learner tries to find a single best action in the current state. It involves learning to search for best actions and trial-and-error. 

One of the promising deep RL methods to derive a good, stable, and easy to use policy is proximal policy optimization (PPO)~\cite{schulman2017proximal}. PPO computes a gradient update at each step that minimizes the cost function while ensuring the deviation from the previous policy is relatively small.  

There are multiple ways to predict the VF and IF from the code embedding. It is possible to use supervised learning methods for example. This, however, would require knowing the labels, \textit{i.e.}, optimal VF and IF for every input loop embedding. To find these labels, it is necessary to run a brute-force search on all the possible VFs and IFs. This might work but can be impractical for a large number of samples. To overcome this challenge we use RL. What distinguishes RL from other machine learning approaches is the presence of self-exploration and exploitation, and the tradeoff between them~\cite{sutton2018reinforcement}. In our case, RL can learn with fewer samples than that required in the supervised learning methods and can co-optimize for multiple objectives such as compilation time, code size, and execution time. 


\section{The Proposed Framework for Automatic Vectorization}
\label{framework}

The proposed framework for automatic vectorization with deep RL and its components are illustrated in Figure~\ref{fig:framework}. The directory of code files is fed to the framework as text code. This code is fed to an automatic loop extractor. The extractor finds and outputs all the loops and their contexts in all the source codes. These outputs are fed to a code embedding generator to learn and generate an embedding. The latter is fed to the deep RL agent to predict the vectorization factors. The agent automatically injects vectorization pragmas as shown in Figure~\ref{fig:action}. The agent then compiles the program with clang/LLVM to gather the execution time improvements, which are used as rewards to the RL agent. Once the model is trained it can be plugged in as-is for inference without further retraining\footnote{It can still be beneficial to keep online training activated so that when completely new loops are observed, the agent can learn how to optimize them too.}.  Note that our framework cannot introduce new errors in the compiled code. Our framework injects pragmas only. These pragmas are used as hints to make vectorization decisions on the loops. However, sometimes the compiler can decide not to consider these pragmas if it is not feasible. For example, predicates and memory dependency can hinder reaching high VF and IF. In that case, if the agent accidentally injected bad pragmas, the compiler will ignore it. 

It is also possible to vectorize from the command line by giving the passes \textit{-force-vector-width=VF} and \textit{-force-vector-interleave=IF}. However, we do not use this option as it restricts us to use a single VF and IF pair for the entire code, which is far from being optimal. Furthermore, the pragma is injected for the most inner loop in the case of nested loops. 
Next, we discuss the details of each component in the proposed framework.

\subsection{Code Embedding}
The ultimate goal of the code embedding generator is to learn a function that maps the input loop codes to a point in a latent multidimensional space where similar loop codes are mapped to points close to each other in the latent multidimensional space. This can allow the RL agent to make similar vectorization decisions on similar codes using the learned embedding. There are multiple ways to generate/learn an embedding for the input code. One example is to use Polly's mathematical representation of loops as an embedding. We see this as a potential future direction for this work. Another example is to use a neural network model pretrained with labels that describe the functionality, \textit{e.g.}, matrix multiplications, dot product, convolution, etc.

In this work we use code2vec~\cite{alon2019code2vec}. Code2vec is a neural network model that relies on natural language processing~\cite{collobert2008unified} and attention~\cite{xu2015show} for representing snippets of code as continuously distributed vectors. Code2vec represents a code snippet as a single fixed-length code vector, which can be used to
predict the semantic properties of the snippet. This vector is composed of $340$ features that embed the program code based on the mapping the code2vec neural network learned. This vector captures many characteristics of the code, such as semantic similarities, combinations, and analogies. The code is first decomposed to a collection of paths in its abstract syntax tree. Then, the network simultaneously learns the atomic representation of each path while learning how to aggregate a set of them. 

\subsection{The RL Environment Definition}
\label{rlenv}

To learn a good policy, it is necessary to appropriately define actions, rewards, and states. 
We define the agent's reward as follows:
\begin{equation}
    reward = (t_{baseline} - t_{RL})/t_{baseline},
\end{equation}
where $t_{baseline}$ is the execution time when compiled with the currently implemented baseline cost model in LLVM and $t_{RL}$ is the execution time when compiled with the injected pragmas by the RL agent. We normalize the execution time by $t_{baseline}$ so that our reward metric is robust to the variations in the programs' execution times. We also use $t_{baseline}$ as a bias in our reward so that a positive reward means the current configuration improves over the baseline. This also reduces the variance in the learned policy.

An action picks the VF and the IF, respectively, from the following values:
\begin{align}
\begin{split}
    VF \in [2^0, 2^1, 2^2, ..., \mathrm{MAX\_VF}],\\
    IF \in [2^0, 2^1, 2^2, ..., \mathrm{MAX\_IF}],
\end{split}
\end{align}
where $\mathrm{MAX\_VF}$ and $\mathrm{MAX\_IF}$ are respectively the maximum VF and IF supported by the underlying architecture. Note that the actions for VF and IF can be defined to have values that are not powers of two. Here they were defined as powers of two only because this is what LLVM currently supports. Initially, we trained two agents, one that predicts VF and the other predicts IF independently. However, from our experiment combining these two agents into one agent with a single neural network that predicts the VF and IF simultaneously performed better. This also aligns with the fact that IF and VF are directly correlated, and in the LLVM compiler code they are defined as a function of each other.

The states of the RL agent were defined as the vector output embedding from the code embedding generator. 
For the inputs of the code embedding generator, we experimented with different snippets of the loop bodies and observed that for nested loops, feeding the loop body of the outermost loop, which also includes the bodies of the inner loops, performed better than feeding the body of the most inner loop only. This is mainly because the entire loop nest better captures the functionally of the code, and reveals the access patterns and strides.

\subsection{Dataset Description}
Neural networks require many samples for training. We first tried to train our model with long-running benchmarks that include code that is not restricted to loops only. It took a long time to train since for every pragma injected for a loop the whole program has to be recompiled and executed. Even if we could overcome the challenge of long execution time with enough resources, the number of open-source benchmarks available for training is very small~\cite{cummins2017synthesizing}.

To speed up the training, and make it more efficient, we built a dataset that includes loops only. We built generators that generate more than 10,000 synthetic loop examples automatically from the LLVM vectorization test-suite. For example, some new tests are made by changing the names of the parameters, which was crucial for reducing noise in the code embedding generator as often the names of the parameters might bias the embedding. Other examples included the stride, the number of iterations, the functionality, the instructions, and the number of nested loops.
 Below are some of the loop examples in the dataset and the (commented) pragma line that the RL agent will inject:
\begin{minted}[fontsize=\footnotesize]{c}
/* Example #1 */
//#pragma clang loop vectorize_width(VF) interleave_count(IF)
for (i = 0; i < N-1; i+=2) {
     assign1[i] = (int) short_a[i];
     assign1[i+1] = (int) short_a[i+1];
     assign2[i] = (int) short_b[i];
     assign2[i+1] = (int) short_b[i+1];
     assign3[i] = (int) short_c[i];
    assign3[i+1] = (int) short_c[i+1];
}
/* Example #2 */
for (i=0; i<M; i++) {
//#pragma clang loop vectorize_width(VF) interleave_count(IF)
    for (j=0; j<N; j++) {
       G[i][j] = x;
     }
}
/* Example #3 */
//#pragma clang loop vectorize_width(VF) interleave_count(IF)
for (i=0; i<N*2; i++){
    int j = a[i];
    b[i] = (j > MAX ? MAX : 0);
}
/* Example #4 */
for (i = 0; i < M; i++){
    for (j = 0; j < L; j++){
        float sum = 0;
//#pragma clang loop vectorize_width(VF) interleave_count(IF)
        for (k = 0; k < N; k++) {
            sum += alpha*A[i][k] * B[k][j];
        }
        C[i][j] = sum;
    }
}
/* Example #5 */
//#pragma clang loop vectorize_width(VF) interleave_count(IF)
for (i = 0; i < N/2-1; i++){
    a[i] = b[2*i+1] * c[2*i+1] - b[2*i] * c[2*i];
    d[i] = b[2*i] * c[2*i+1] + b[2*i+1] * c[2*i];
}
\end{minted}
\begin{figure}[!t]
    \centering
    \includegraphics[width=0.45\textwidth]{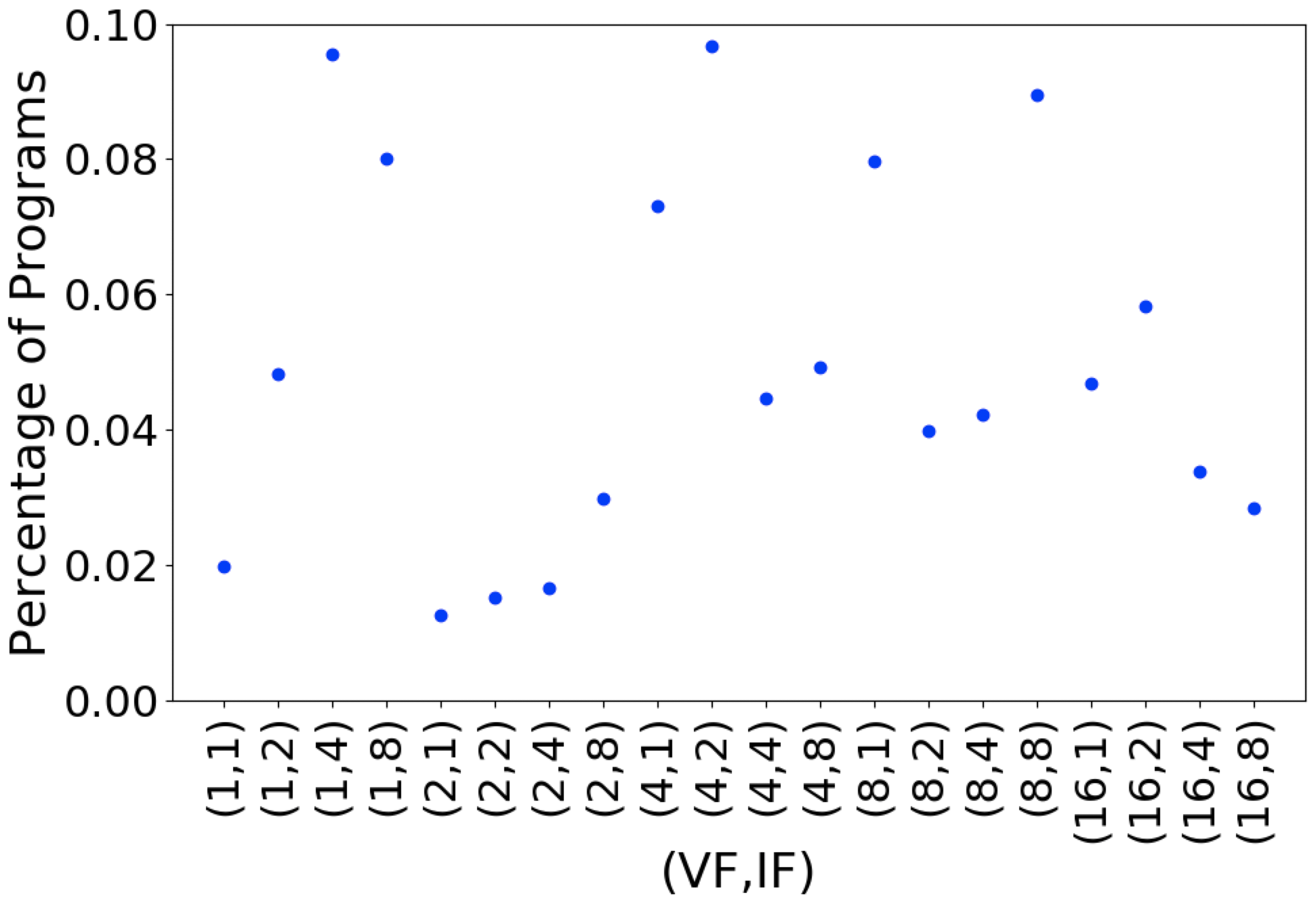}
    \caption{The distribution of optimal VF and IF with brute-force search for different programs in the dataset.}
    \label{fig:percentage}
\end{figure}
Figure~\ref{fig:percentage} shows the distribution of optimal vectorization factors when running a brute-force search with  $\mathrm{MAX\_VF}=16$ and $\mathrm{MAX\_IF}=8$ on the dataset. While these loops do not represent all the existing loops, the results show that different loops have different optimal VF and IF and to guarantee optimal performance, all combinations of VF and IF should be considered. Interestingly, the factors with the highest percentage of programs are $(VF=4, IF=2)$. From our experiments, these factors were the default values the baseline cost model also outputted.

\subsection{Handling Long Compilation Time}
During training, some of the programs took a long time to compile, mainly when the agent was trying to vectorize more than plausible. Long compilation time with limited resources can slow down the training. To overcome this, we limited the compilation time to ten times the time it takes to compile a program with the baseline cost model. If the program took longer than that to compile, we gave a penalty reward of $-9$ (equivalent to assuming it takes ten times the execution time of the baseline) so that the agent will learn not to overestimate the vectorization and avoid it. From our experiments on the programs that took a relatively long time to compile, eventually after waiting the necessary time for them to compile, the achieved performance was not better than that of all the other possible vectorization configurations. In some contexts, users might care about compile-time when evaluating the performance of programs. Our reward definition can incorporate that too so that the agent can simultaneously optimize for more than one objective.  For example, one can allow a long compilation time but penalize for it. The reward can also be defined as a combination of the compilation time, execution time, generated assembly code size, \textit{etc}. 



\section{Evaluation}
\label{eval}
\begin{figure*}[!t]
    \centering
    \includegraphics[trim={0cm 0cm 0cm 0.05cm},clip,width=\textwidth]{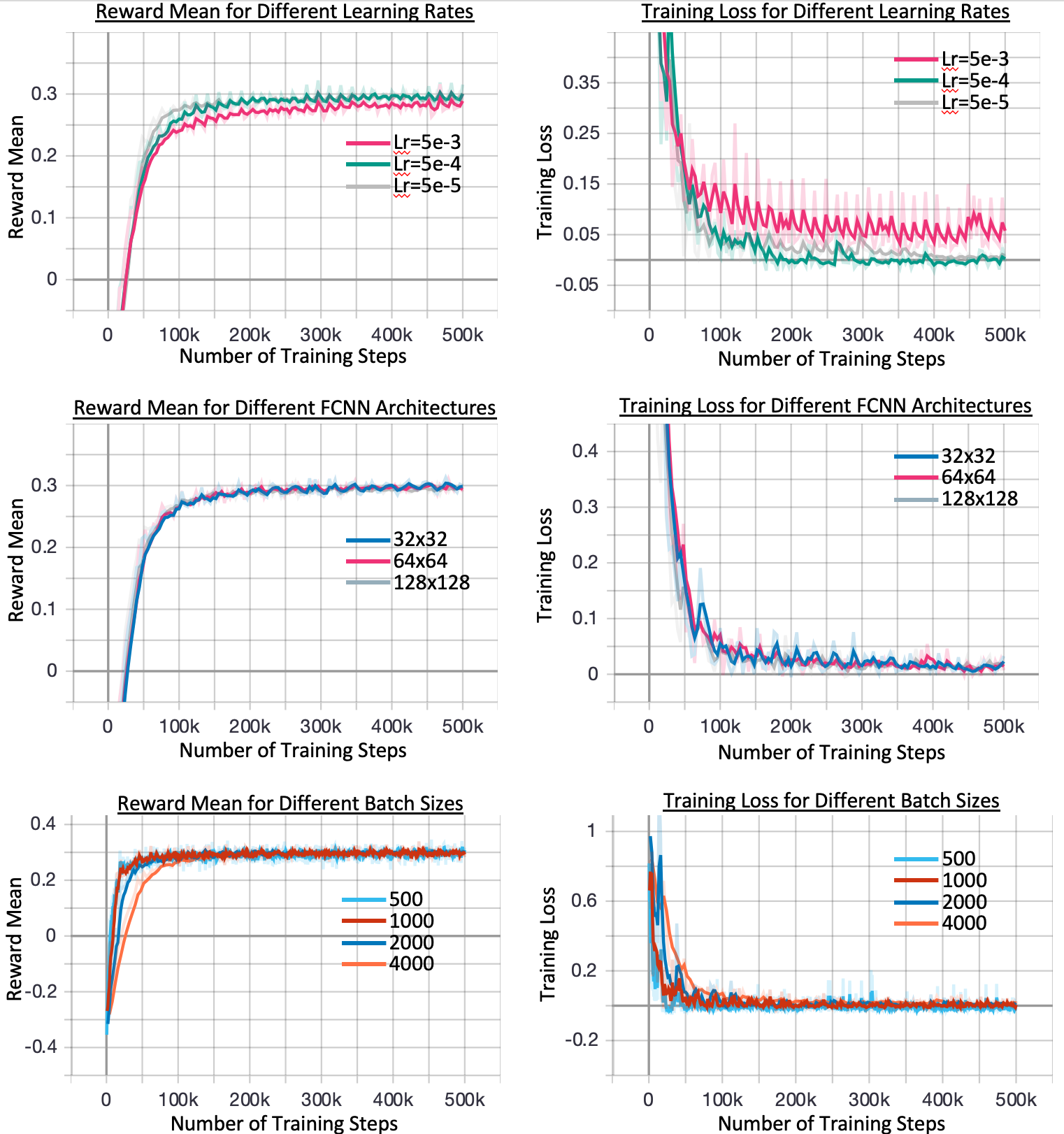}
    \caption{Reward mean and training loss for different learning rates, FCNN architectures, and batch sizes.}
    \label{fig:lc}
\end{figure*}

The proposed framework is evaluated following the methodology mentioned in Section~\ref{motivation}. For code2vec we use the open-source code and modify it to work with our RL agent implementation. To run our RL algorithms we use RLlib~\cite{liang2017ray} and Tune~\cite{liaw2018tune}, open-source libraries for RL that offer, high scalability, hyper-parameter tuning and a unified API for a variety of applications. RLlib and Tune are built on top of Ray~\cite{moritz2018ray}, a high-performance distributed execution framework targeted at large-scale machine learning and RL applications.  We first train the framework with the RL agent and code2vec until convergence. Then we also run a brute-force search on the dataset to find the best vectorization factors and use them as labels for NNS, the decision tree and the supervised FCNN. Since the brute-force search requires a long time to run, we limit our training set to 5,000 samples and use this set for the rest of our evaluation. To report performance we take twelve completely different benchmarks from the test set. These benchmarks combine completely different benchmarks from the LLVM test-suite. These benchmarks include loops with different functionality and access patterns. For example, predicates, memory accesses with different strides, bitwise operations, unknown loop bounds, if statements, unknown misalignment, multidimensional arrays, summation reduction, type conversions, different data types, \textit{etc}.  We compare the performance of our framework versus Polly and the baseline cost model.
\begin{figure*}
    \centering
    \includegraphics[trim={0cm 0cm 0cm 1.4cm},clip,width=\textwidth]{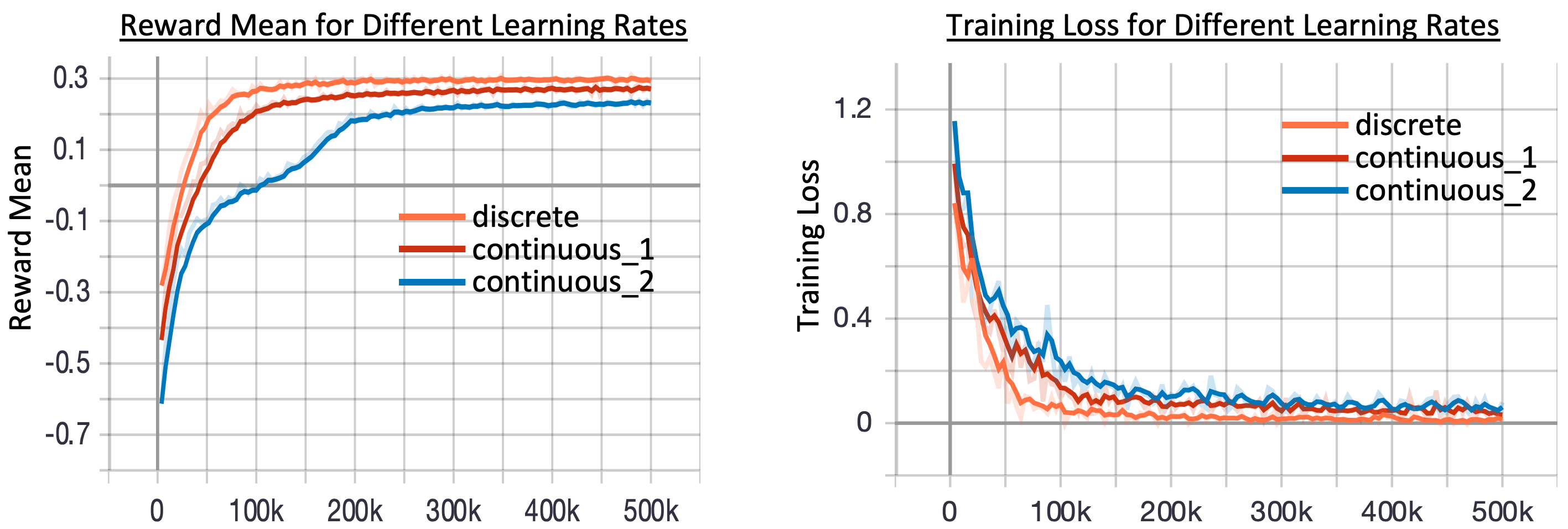}
    \caption{Reward mean and training loss for different action space definitions.}
    \label{fig:lc2}
\end{figure*}

We start with a $64\times 64$ FCNN, with training batch size of 4,000, a learning rate of 5e-5 - a hyperparameter which determines to what extent newly acquired information overrides old information - and discrete actions. We then experiment with changing one parameter at a time. For discrete actions, the neural network picks two integer numbers that index into the arrays of possible VFs and IFs. We experiment with different hyperparameters. Figure~\ref{fig:lc} shows a hyperparameter sweep over different hyperparameters as function of number of training steps, \textit{i.e.}, compilations. We train up to 500,000 steps to see whether more training can get to better rewards, but it is clear the policy converges with much fewer steps.

These results show that the current framework is robust to noise and different parameters. When the learning rate was set to 5e-5 the reward mean reached the maximum the fastest. For learning rate 5e-3 the reward mean never reached a higher maximum than that of the smaller learning rates and the training loss was the highest. Minor differences were observed for the different FCNN architectures. We also tried single hidden layer networks and deeper networks and the results were similar so they were not included in the figures for clarity. The policy converged with fewer samples as the batch size was decreased. We also experimented with smaller batch sizes and they resulted in unstable policies that did not outperform the performance when the batch size was set to 500.

The results also show that the policy converged and arrived at a highly rewarding (higher than 0 means better than the baseline on average based on the reward definition described in Section~\ref{rlenv}) state with 5,000 samples (for the lowest batch size); $35\times$ less than that required for a brute-force search or a supervised learning method. It is important to point out that this training is performed once and later the framework can be used for inference, which requires a single step only, similar to the baseline cost model. On the other hand, a brute-force search would require searching again.

Figure~\ref{fig:lc2} shows the reward mean and total training loss as function of number of training steps for different action space definitions. We experimented with three actions space definitions: \circled{1} discrete action space where the agent picks two integer numbers that correspond to indices in the arrays of VFs and IFs. \circled{2} Continuous action space where the agent picks one continuous number that encodes both the VF and IF. \circled{3} Continuous action space where the agent picks two continuous numbers that encode both the VF and IF. The numbers in the continuous action spaces are rounded to the closest integers. The results show that the discrete action space performs the best.

\begin{figure*}
    \centering
    \includegraphics[trim={0cm 0cm 0cm 0cm},clip,width=\textwidth]{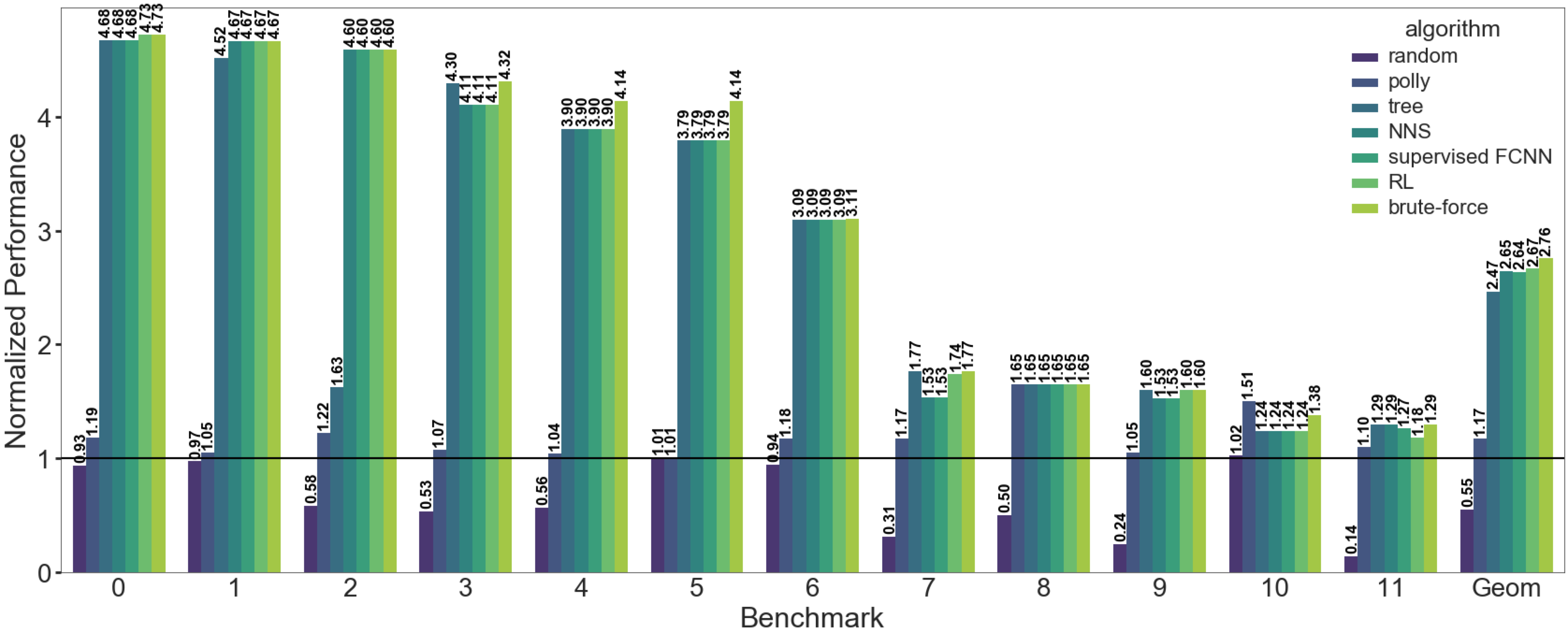}
    \caption{The performance of the proposed vectorizer that can be configured to use NNS, random search, decision trees, and RL compared to brute-force search, Polly and the baseline cost model. The performance is normalized to the baseline.}
    \label{fig:myperf}
\end{figure*}

The performance on different benchmarks for the baseline, random search, Polly, decision tree, NNS, supervised FCNN, and RL and brute-force search are shown in Figure~\ref{fig:myperf}. RL outperformed the baseline by $2.67\times$ on average and achieved performance only $3\%$ worse than that of the brute-force search. The performance differed between the different benchmarks based on how much vectorization the program can absorb. NNS and decision trees also performed well, achieving respectively $2.65\times$ and $2.47\times$ better than the baseline. This shows that the embedding learned by the code embedding generator during the end-to-end training is good so that other learning methods that cannot be trained end-to-end can use this embedding and perform well.

Random search performed much worse than the baseline. This shows that the framework learned a structure in the observations that manifested in the vectorization decisions it made. Polly outperformed the baseline by $17\%$ but performed $56\%$ worse than the proposed RL solution. For benchmark \#10, Polly interestingly outperforms the brute-force search. This is because Polly performs loop transformations that optimize beyond vectorization. This also shows the potential for achieving better performance improvement when combining Polly and deep RL. We plan to explore this option in future work.

\begin{figure}[!h]
    \centering
    \includegraphics[ width=0.47\textwidth]{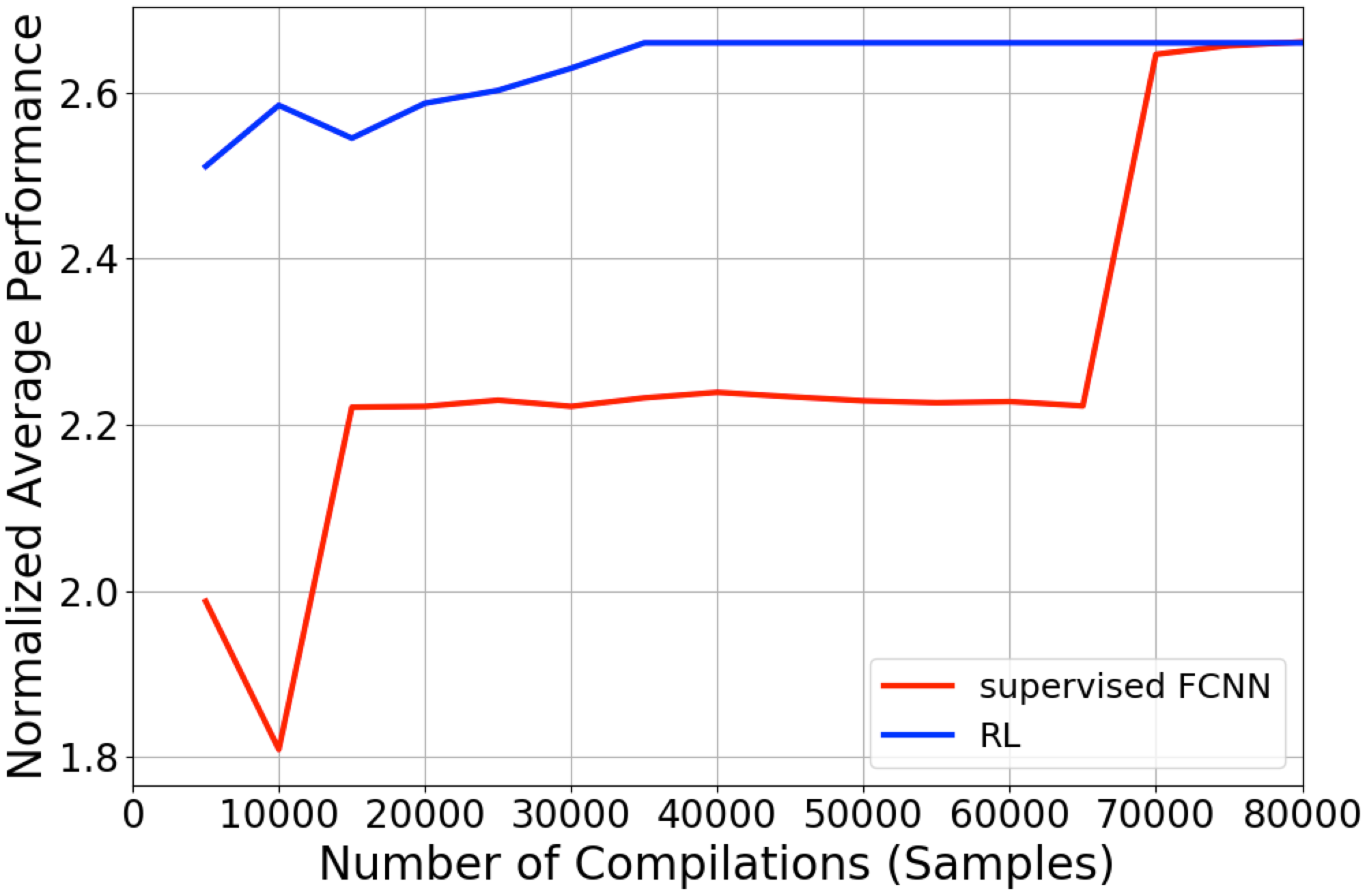}
    \caption{Normalized average performance of supervised FCNN and deep RL as a function of the number of compilations used (samples) for training.}
    \label{fig:efficiency}
\end{figure}

While NNs and decision trees cannot be trained end-to-end and require special handling, the supervised FCNN can be trained end-to-end and achieves comparable performance to deep RL. However, RL does not require labels and thus can be trained without a brute-force search. To demonstrate the advantage of deep RL, Figure~\ref{fig:efficiency} shows the normalized average (geomean) performance of deep RL compared to supervised FCNN as a function of the number of compilations required (samples). Deep RL with as low as 5,000 compilations already achieves high performance (only $5\%$ worse than the peak) and $1.26\times$ better than the supervised FCNN. Supervised FCNN achieved this only after 70,000 compilations making it $14\times$ less sample efficient than deep RL. Furthermore, in the long run we believe deep RL can better handle large search spaces with multiple objectives to co-optimize.

\begin{figure}
    \centering
    \includegraphics[width=0.5\textwidth]{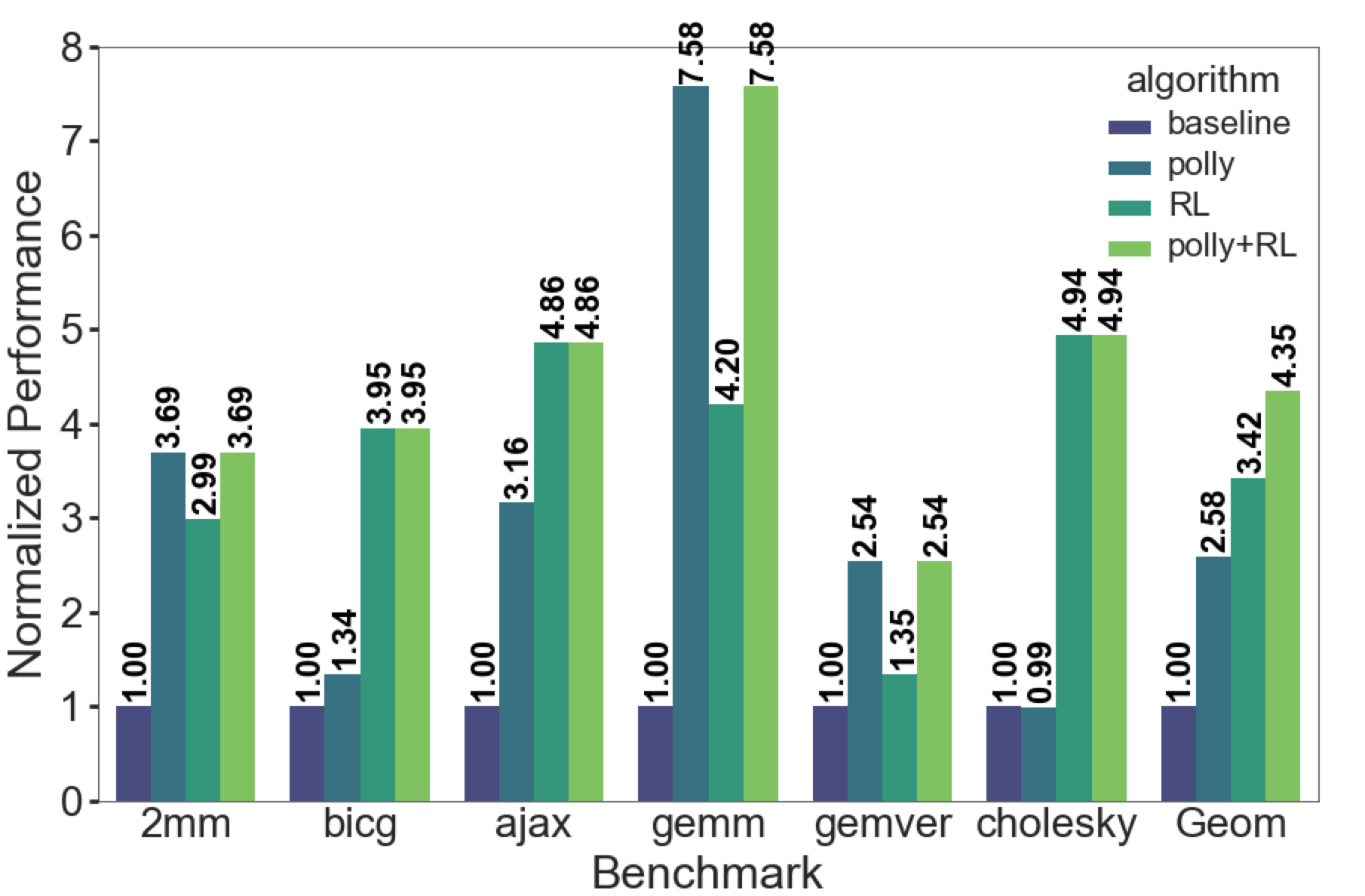}
    \caption{The performance of the proposed vectorizer on Polybench compared to Polly and the baseline cost model. The performance is normalized to the baseline.}
    \label{fig:myperf3}
\end{figure}

\subsection{Transfer Learning}
The goal of this subsection is to see how well the framework generalizes to a completely new code. To that end we evaluate the trained model on two benchmarks: MiBench~\cite{guthaus2001mibench} where the loops constitute a minor portion of the code and PolyBench~\cite{pouchet2012polybench} where the loops constitute a major portion of the code. MiBench is a set of free and commercially representative embedded benchmarks such as telecommunication, networking, security, office, and automation. Note that vectorization for some of the MiBench benchmarks is not possible. For example, due to memory dependencies, control-flow or lack of loops, it was not possible to vectorize \textit{adpcm, dijkstra, basicmath, blowfish, etc.} PolyBench includes benchmarks that perform matrix operations, decomposition, and linear algebra for which Polly is optimized to run on. 

Figure~\ref{fig:myperf3} shows the performance of deep RL, Polly and the baseline on PolyBench. Deep RL achieves on average $3.42\times$\footnote{Note that we take the average performance improvement over multiple inferences. If instead we take the best performance, the improvement reaches $4.77\times$ on average: $3.71\times$, $6.74\times$, $6.92\times$, $5.21\times$, $1.61\times$, and $8.16\times$ for \textit{2mm, bicg, ajax, gemm, gemver,} and \textit{cholesky}, respectively} better performance than the baseline and $1.33\times$ better than Polly. Polly was optimized to run on PolyBench, yet deep RL outperformed Polly on three out of the six benchmarks. From deeply investigating the different benchmarks we found that Polly performed better on some benchmarks due to the ability of Polly to perform loop transformations that optimize beyond vectorization, to lack of enough benchmarks in the dataset to fully represent the space of loops, and to the high penalty we give to long compilation times. When the deep RL agent tried to give high VF and IF the reward sometimes decreased due to the high penalty we give to long compilation times. In such cases, the agent learns to avoid being over-optimistic about increasing the VF and IF. With more training data the agent can generalize better to larger loop bounds on new examples. When combining Polly and deep RL the average performance improvement that can be achieved (potentially) is $4.35\times$. 

\begin{figure}
    \centering
    \includegraphics[width=0.5\textwidth]{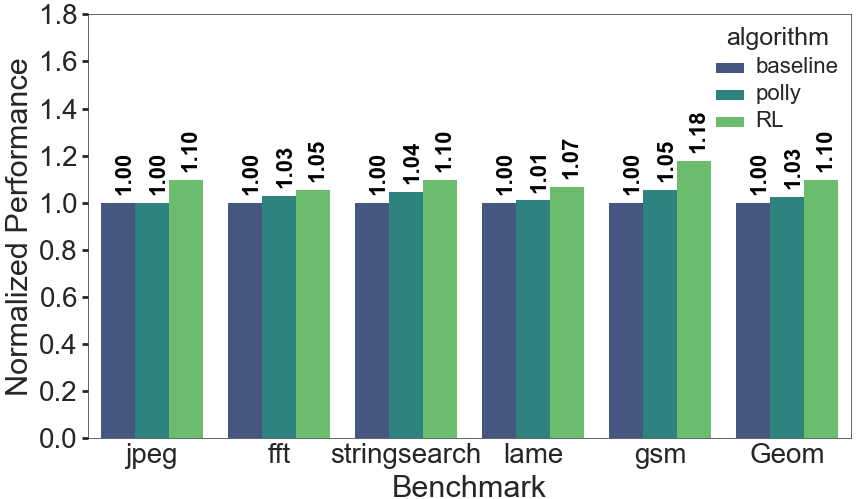}
    \caption{The performance of the proposed vectorizer on Mibench compared to Polly and the baseline cost model. The performance is normalized to the baseline.}
    \label{fig:myperf2}
\end{figure}

Figure~\ref{fig:myperf2} shows the performance of deep RL, Polly and the baseline on MiBench. Deep RL outperforms both Polly and the baseline in all the benchmarks. The average performance improvement was $1.1\times$ over the baseline. While this might not seem considerable, we believe that it can be sufficient since the benchmarks did not rely heavily on loops, and the measured execution time was for all the code not restricted to loops.

\subsection{Discussion: Deployability}
In general, vendors and commercial companies are reluctant to adopt machine learning and deep learning methods in compiler optimization. The main reason behind this is the need for methods that are deterministic, simple, easy to explain, and performant on a large scale of applications. This also explains why most of the optimizations and implementation in compilers are based on manual engineering and heuristics. With that being said, we believe that the growing complexity in systems and workloads, and availability of data demands learning-based approaches. Deep RL and other deep learning methods present a unique opportunity to address these compiler challenges end-to-end and improve upon manual engineering. In our evaluation, we showed that deep RL can generalize to new benchmarks. With enough training data, deep RL can be deterministic and performant on a large scale of applications. Since the use of deep RL will mainly be for inference it will also be simple to use and deploy. The main challenge will remain in interpretability. This challenge is not only a limitation of deep RL in vectorization, it is also a limitation of neural networks in general. Many recent works are being conducted on explaining neural network decisions~\cite{gunning2017explainable} and their application in code optimization will also benefit from that. Besides, neural networks have been adopted to solve many advanced real-world challenges regardless of the interpretability limitation. We believe that compilers and code optimization should also follow.

\section{Future Directions}
\label{future}
We see multiple future directions for this work. It is possible to use loop polyhedral analysis, which is dedicated to the loop snippets of codes for the code embedding. This will also be less expensive in terms of computations. Combining deep RL and Polly can further boost the performance and the RL agent can also be trained to predict whether to use Polly or not. The deep RL vectorizer can also be employed at the intermediate representation level, which can better reflect the effects of the vectorization on the code and thus could enable learning better predictions. For different target architectures, it is necessary to add features that represent the underlying architecture or to train separate models that are fitted to the used architecture as different architectures behave differently and have different VF and IF action spaces. In this work, we showed the potential of the deep RL vectorizer as the first step toward end-to-end code optimization with machine learning and deep RL. It is, however, necessary to train on a wide range of applications, and target architectures for the deep RL vectorizer to be a standardized optimization stage in the LLVM compilation stack. 

In our approach, we assumed the agent makes a single decision per loop nest (\textit{i.e.}, episode). However, with new compiler features such as the support of vectorization at different levels of a nested loop, deep RL will be more attractive. This is mainly because the deep RL agent is not restricted to make a single decision per loop nest. Instead, it can perform multiple sequential decisions that collectively form a single episode of multiple actions and states.

Pragmas like function inlining, loop unrolling, superword-level parallelism, and scatter/gather can also be tuned in a similar manner. The user only needs to define an appropriate action space and a reward function that depends on the desired objective. Many of the optimizations done today in the compiler are global rather than local. For example, the phase ordering of compiler passes is applied drastically to all the functions in the code. It can be possible to automatically determine different phase orderings and optimizations to different sections of the code. 

Our framework can also support vanilla deep neural networks methods instead of deep RL. One direction we are exploring is to use a neural network that learns a ranking scheme on the VF and IF. For example, it can learn that given an embedding, and pragmas, what will the execution time normalized to the non-vectorized code be. This is equivalent to learning a new cost model for the different VFs and IFs, which could potentially replace the baseline cost model used today. This method - unlike NNs and decision trees - can be trained end-to-end. 
\section{Related Work}
\label{related}
Previous work has utilized machine learning in compiler optimization~\cite{ashouri2018survey,wang2018machine}. 
For example, the work in~\cite{haj2019autophase,fursin2011milepost} proposed deep supervised and RL methods to overcome the phase ordering challenge. In~\cite{stock2012using}, multiple machine learning methods for automatic vectorization have been proposed. Our work is different from these prior works in that it is the first to propose a solution based on deep RL to explore the vectorization space and compiler optimization in general. Second, all these works primarily rely on extracted/engineered (hand-crafted) features from the program, e.g., arithmetic intensity, memory operations, number of different instruction, distance between producer and consumer, etc. These features however do not fully represent the original code. By contrast, our work addresses the automatic vectorization by learning useful features in an end-to-end fashion, from the text code itself to the optimal factors without any loss of information. In~\cite{cummins2017end} end-to-end supervised deep learning is used to learn compiler heuristics. While such approach can achieve comparable performance, finding the labels for training can be time consuming, and optimizing for multiple objectives with large search spaces can be challenging. 

Automatic vectorization with other methods was also proposed. For example, the currently implemented cost model in LLVM and recently proposed cost models in~\cite{tian2016llvm,trifunovic2009polyhedral,nuzman2011vapor} rely on predefined cost functions that calculate the expected execution time of a vectorized loop based on a linear formula from the instruction distribution. \cite{porpodas2015throttling} improves super-word level parallelism (SLP)~\cite{larsen2000exploiting} to limit the automatic vectorization. This work does not address loop vectorization and relies on the baseline cost model to predict when some portions of code are better off not vectorized. Also, \cite{mcfarlin2011automatic} relies on heuristics to automatically vectorize. Finally, \cite{porpodas2017supergraph,porpodas2015pslp} improve SLP and rely on fixed cost models such as weighted instruction count or the current LLVM cost models.
\section{Conclusion}
\label{conc}
In this work, we proposed and developed an end-to-end vectorization framework that automatically detects loops, learns their structures and applies deep RL to inject vectorization pragmas to the compiler. Our results demonstrated an average performance improvement $1.29\times-4.73\times$ compared to the baseline cost model implemented in LLVM and on average only $3\%$ worse than the brute-force solution. Looking forward, we foresee a potential opportunity for automatic end-to-end code tuning and optimization with machine learning techniques, such as deep RL.
\begin{acks}
The authors would like to thank Ronny Ronen, Ayal Zaks, Gadi Haber, Hideki Saito, Pankaj Chawla, Andrew Kaylor and anonymous reviewers for their insightful feedback and suggestions. 
\end{acks}
\bibliography{main}

\end{document}